\documentstyle[psfig]{mn}

\newif\ifAMStwofonts

\title[Spectral ageing analysis of GRSs]
      {A multifrequency study of giant radio sources \\ II. Spectral ageing analysis of the lobes
of selected sources}
\author[M. Jamrozy et al.]
        {M. Jamrozy$^1$$\thanks{E-mail: jamrozy@oa.uj.edu.pl (MJ); chiranjib@iucaa.ernet.in (CK); 
       machalsk@oa.uj.edu.pl (JM); djs@ncra.tifr.res.in (DJS)}$, C. Konar$^{2,3}$, J. Machalski$^1$
        and D.J. Saikia$^{2}$ \\ 
$^1$ Obserwatorium Astronomiczne, Uniwersytet Jagiello\'nski, ul. Orla 171, 30244 Krak\'ow, Poland \\
$^2$ National Centre for Radio Astrophysics, TIFR, Pune University Campus, Post Bag 3, Pune 411 007, India\\ 
$^3$ Inter-University Centre for Astronomy and Astrophysics, Pune University Campus, Post Bag 4, 
     Pune 411 007, India }
\date{Accepted.    Received }

\pagerange{\pageref{firstpage}--\pageref{lastpage}}

\begin{document}

\maketitle

\label{firstpage}

\begin{abstract}
Multifrequency observations with the Giant Metrewave Radio Telescope (GMRT) 
and the Very Large Array (VLA) are used to determine
the spectral breaks in consecutive strips along the lobes of a sample of
selected giant radio sources (GRSs) in order to estimate their spectral ages. 
The maximum spectral ages estimated for the detected radio emission in the lobes
of our sources range from $\sim$6 to 36 Myr with a median value of $\sim$20 Myr
using the classical equipartition fields. Using the magnetic field estimates from
the Beck \& Krause formalism the spectral ages range from $\sim$5 to 38 Myr with 
a median value of $\sim$22 Myr. These ages are significantly older than smaller sources. 
In all but one source (J1313+6937) the spectral age gradually increases
with distance from the hotspot regions, confirming that acceleration of the
particles mainly occurs in the hotspots. Most of the GRSs do not exhibit zero
spectral ages in the hotspots, as is the case in earlier studies  of smaller
sources. This is likely to be largely due to  contamination by more extended emission 
due to relatively modest resolutions. The injection spectral indices range from $\sim$0.55 to 
0.88 with a median value of $\sim$0.6. We discuss these values in the light of
theoretical expectations, and show that the injection spectral  index appears to be
correlated with luminosity and/or redshift as well as with linear size.  
\end{abstract}


\section{Introduction}
The radio continuum spectra in different parts
of an extended radio source contain important information about 
the various energy losses and gains of the radiating
particles during the lifetime of the source. According to dynamical models 
for FRII-type (Fanaroff \& Riley
1974) radio sources, the energy emitted by the nucleus squirts out in the form of 
narrow, collimated jets channeling their way through the external environment. These jets
dissipate their energy at their leading edges
giving rise to intense regions of emission called `hotspots'. As the jets
advance outwards, the relativistic particles flow out from the hotspots to form the
extended lobes of radio emission.  Assuming that there is no significant
reacceleration within these lobes and no significant mixing of particles, there
should be a spectral gradient across the radio source, in the sense that the spectrum
should steepen with increasing distance from the hotspot. This prediction has been seen in 
many sources and used to estimate the radiative ages and expansion
velocities in several samples of powerful 3CR sources (e.g. Myers \& Spangler 1985;
Alexander \& Leahy 1987; Leahy, Muxlow \& Stephens 1989; Carilli et al. 1991; Liu, Pooley \& Riley 1992),
in samples of low-luminosity and medium-luminosity radio galaxies (e.g. Klein et al.
1995; Parma et al. 1999), as well as a sample of compact steep-spectrum sources
(Murgia et al. 1999). However, the observed spectral break and steepening of the
spectrum beyond this break need not be entirely due to radiative energy losses. 
A possible role of the local magnetic fields, details of the backflow of
the lobe material, or the difficulties in disentangling various energy losses of
the radiating particles have been pointed out in a number of papers (e.g. Wiita \&
Gopal-Krishna 1990; Rudnick,
Katz-Stone \& Anderson 1994; Eilek \& Arendt 1996; Jones, Ryu \& Engel 1999; 
Blundell \& Rawlings 2000).

\begin{figure*}
\vspace{130mm}
\includegraphics{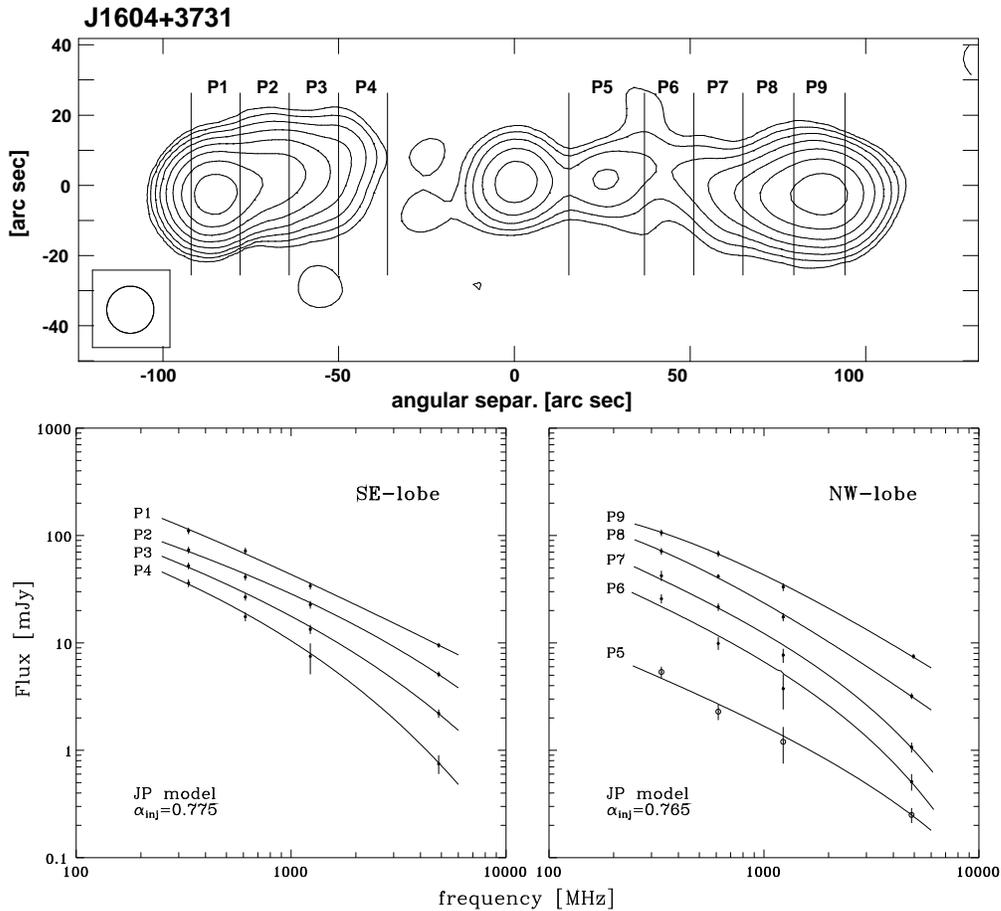}
\caption{Upper panel: an example of how the lobes of analysed sources are cut into
strips within which the spectral age of radiating particles is determined.
Lower panels: the spectra fitted to the flux-density data in each of the strips.}
\end{figure*}

Nevertheless, giant-sized radio sources (hereinafter referred to as GRSs) are 
suitable for the classical spectral-ageing analysis due to their large angular extent
which can be covered by a significant number of resolution elements. In Paper I
(Konar et al. 2007) extensive low-frequency 
observations made with the GMRT of a sample of ten selected GRSs
have been described and some of their properties analysed and discussed.  
In this paper we analyse the spectral ages of these radio galaxies
except for J0720+2837 for which we have data at only two frequencies and J1343+3758
whose spectral ageing analysis has been presented by Jamrozy et al. (2005). Our analysis
is similar to the spectral ageing analysis for the GRS J1343+3758 (Jamrozy
et al. 2005) and the double-double radio galaxy J1453+3308 (Konar et al. 2006).
In Section 2 we discuss briefly the method of analysis, while in Section 3 we present 
our results.  Discussion of the results and an examination of the correlations of the
injection spectral index, $\alpha_{\rm inj}$, with luminosity, redshift and linear size
are presented in Section 4. The conclusions are summarized in Section 5.

\section{Spectral ageing analysis}

\subsection{Spectral steepening and spectral age}
The spectral age, i.e. the time which has elapsed since the particles were last
accelerated in the hotspot region, can be derived from the steepening
in the radio spectrum due to energy losses caused by synchrotron and 
inverse-Compton processes. The observed spectra have been fitted using the
Jaffe \& Perola (1973, JP) and the Kardashev-Pacholczyk (Kardashev 1962;
Pacholczyk 1970; KP) models using the {\tt SYNAGE} package (Murgia 1996).
We found no significant difference between these two models over the frequency range
of our observations and present only the results obtained from the JP model in this paper.  
Also, while fitting the theoretical spectra to the observed ones, we find the spectra 
of the different strips in the lobes are better fitted with the JP model than with 
the continuous injection (Kardashev 1962; CI) model, though sometimes the CI model 
fits better in the area of a prominent hotspot.

Thus, under assumptions that (i) the magnetic field strength in a given lobe is
constant throughout the energy-loss process, (ii) the particles injected into the
lobe have a constant power-law energy spectrum with an index $\gamma$, and
(iii) the time-scale of isotropization of the pitch angles of the particles is short 
compared with their radiative lifetime, the spectral age, $\tau_{\rm spec}$, is given by

\begin{equation}
\tau_{\rm spec}=50.3\frac{B^{1/2}}{B^{2}+B^{2}_{\rm iC}}\left\{\nu_{\rm br}(1+z)\right\}
^{-1/2} [{\rm Myr}],
\end{equation}

\noindent
where $B_{\rm iC}$=0.318(1+$z$)$^{2}$ is the magnetic field strength equivalent to
the cosmic microwave background radiation. Here $B$, the magnetic field strength of the
lobes, and $B_{\rm iC}$ are expressed
in units of nT, $\nu_{\rm br}$ is the spectral break frequency in GHz above which the
radio spectrum steepens from the initial power-law spectrum given by
$\alpha_{\rm inj}$=$(\gamma-1)/2$. It may be noted that Alexander \& Leahy (1987) and
Alexander (1987) have shown that the effects of expansion losses may be neglected.

\begin{figure}
\vspace{140mm}
\includegraphics{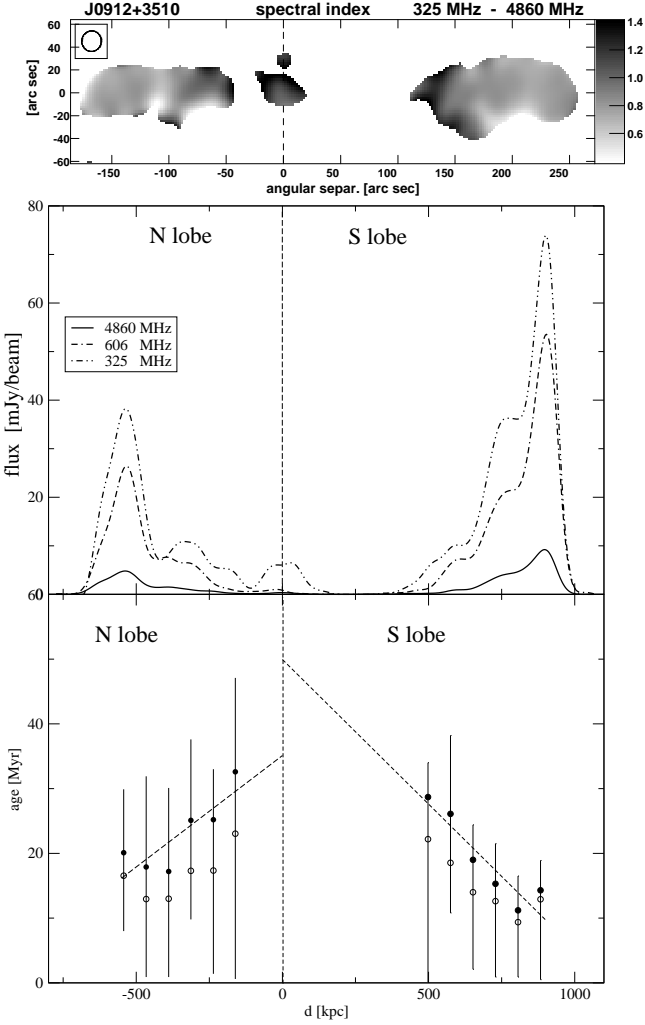}
\caption{Spectral-index map, flux-density profiles, and the spectral age distribution
for J0912+3510. The spectral ages have been estimated using magnetic field values 
determined using the Beck \& Krause (2005) formalism (filled circles and with error bars)
and the classical (e.g. Miley 1980; referred to as classical-1 in Paper I) formalism
(open circles without error bars).}
\end{figure}

\subsection{Determination of $\alpha_{\rm inj}$ and $\nu_{\rm br}$ values} 
In order to determine a value of $\alpha_{\rm inj}$, we fit the JP model 
to the flux densities of the entire lobes (given in Paper I) treating $\alpha_{\rm inj}$ 
as a free parameter. We find that the uncertainties of the $\alpha_{\rm inj}$
values are sometimes large due to the low flux densities of the lobes.
We have used the values of  $\alpha_{\rm inj}$ which give the best fits to the
spectra of the lobes, and note that there is usually no evidence for significantly different 
values of this parameter for the oppositely directed lobes. 
For sources with reliable flux densities at lower frequencies than our observations but with
no information on the flux densities of the individual lobes at these frequencies,
we have determined $\alpha_{\rm inj}$ from the integrated spectra of the sources.
For these cases we have used the same value of $\alpha_{\rm inj}$ for both the lobes.  
Having estimated the $\alpha_{\rm inj}$ values, the total-intensity maps of the 
GRSs published in Paper I are convolved to
a common angular resolution, which corresponds to the lowest resolution of our images. 
Each lobe is then split into a number of strips,
separated approximately by the resolution element along the axis of the 
source in such a way that the extreme strips are centred at the peaks of brightness
on the convolved maps. In each source a region around the radio core is
excluded. As an example, the division of the source J1604+3731 into different strips
is shown in Fig. 1.  Using the {\tt SYNAGE} software we determine the best
fit to the spectrum in each strip over the entire observed frequency range 
using the JP model, and derive the relevant
value of $\nu_{\rm br}$. The fitted spectra in the strips covering the SE-lobe and
the NW-lobe of J1604+3731 are plotted in the lower panels of Fig. 1. 

\begin{table}
\caption{Break frequency, magnetic field strength, and spectral age of particles in
 consecutive strips through the lobes for J0912+3510}
\begin{tabular}{@{}llrlcl}
\hline
Strip & Dist. & $\nu_{\rm br}$ & $\chi^{2}_{\rm red}$ & $B_{\rm eq}$(rev) & $\tau_{\rm spec}$ \\
      & (kpc) &(GHz) &    & (nT) & (Myr) \\
\hline
   &         &    {\bf N-lobe}            &           &$\rm\alpha_{inj}=0.560$&                    \\
P1 & 543     & $12.6^{+32.9}_{-7.7}   $   &0.07       & 0.39$\pm$.032  & $ 20.1^{+9.8}_{-12.1}  $ \\
P2 & 466     & $17.2^{>+85}_{-11.8}   $   &0.02       & 0.31$\pm$.028  & $ 17.9^{+14.0}_{-17.0} $ \\
P3 & 389     & $18.6^{>+90}_{-12.5}   $   &0.35       & 0.31$\pm$.027  & $ 17.2^{+12.9}_{-16.3} $ \\
P4 & 313     & $ 8.8^{+48.0}_{-4.9}   $   &0.08       & 0.29$\pm$.027  & $ 25.1^{+12.5}_{-15.3} $ \\
P5 & 236     & $ 8.6^{>+70}_{-3.5}    $   &0.07       & 0.26$\pm$.025  & $ 25.2^{+7.8}_{-23.8}  $ \\
P6 & 161     & $ 4.9^{>+80}_{-2.5}    $   &0.38       & 0.22$\pm$.022  & $ 32.6^{+14.5}_{-32.0} $ \\
    &         &                            &          &                &                         \\
    &         &    {\bf S-lobe}            &          &$\rm\alpha_{inj}=0.628$&                   \\
P7  & 498     & $ 6.7^{>+90}_{-1.9}      $ &9.93      & 0.28$\pm$.031  & $28.7^{+5.3}_{-28.6}  $ \\
P8  & 575     & $ 8.1^{+38.6}_{-4.3}     $ &0.00      & 0.31$\pm$.034  & $26.1^{+12.1}_{-15.3} $ \\
P9  & 652     & $14.9^{+13.2}_{-10.7}    $ &2.05      & 0.34$\pm$.037  & $19.0^{+5.4}_{-16.9}  $ \\
P10 & 729     & $21.6^{+35.2}_{-15.9}    $ &1.06      & 0.40$\pm$.042  & $15.3^{+6.2}_{-14.4}  $ \\
P11 & 806     & $39.6^{+96.7}_{-31.3}    $ &2.52      & 0.41$\pm$.042  & $11.2^{+5.3}_{-10.3}  $ \\
P12 & 883     & $22.6^{+23.0}_{-16.7}    $ &6.92      & 0.45$\pm$.046  & $14.3^{+4.6}_{-13.8}  $ \\        
\hline
\end{tabular}
\end{table}
 
\subsection{Magnetic-field strength determination}
In order to estimate the spectral ages in different parts of the lobes,  we
have to estimate the magnetic-field strength in the consecutive strips. Following Konar
et al. (2006, 2007), the values of the equipartition energy density and the corresponding
magnetic field, $B_{\rm eq}$, are calculated using the revised formalism proposed by
Beck \& Krause (2005, hereinafter referred to as BK). This has been described in some detail 
in Paper I, and the
values are denoted by $B_{\rm eq}$(rev) in this paper. The revised field
strength within each of the consecutive strips is calculated assuming a filling
factor of unity. Since the BK formalism yields magnetic field strengths 
which are larger by a factor of $\sim$3 compared with the classical formalism
(e.g. Miley 1980; hereinafter refererred to as classical-1, as in Paper I), we also
estimate the spectral ages using the classical-1 magnetic field strengths. This shows
the range of ages possible for different estimates of the magnetic field strength.  
Parameters derived using the latter field strenghts are denoted by `class' in this paper.  
It is relevant to note that in most of the previously published papers on spectral 
ageing analysis of smaller FRII-type radio sources the spectral ages were determined
using the classical minimum-energy formalism. 

\begin{figure}
\vspace{140mm}
\includegraphics{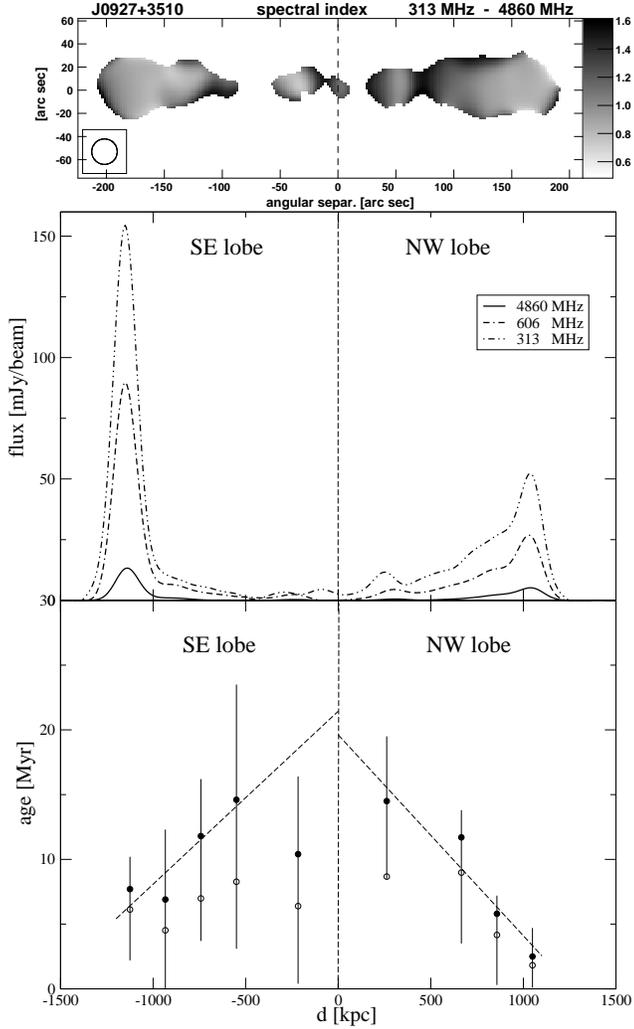}
\caption{As in Fig.\,2, but for J0927+3510}
\end{figure}

\begin{table}
\caption {As in Table 1 but for J0927+3510}
\begin{tabular}{lrrlcl}
\hline
Strip & Dist. & $\nu_{\rm br}$ & $\chi^{2}_{\rm red}$ & $B{\rm eq}$(rev)& $\tau_{\rm spec}$ \\
      & (kpc) &(GHz) &    & (nT) & (Myr) \\
\hline
    &         &    {\bf SE-lobe}                &       &$\rm\alpha_{inj}=0.700$&                \\
P1 & 1125    &     $20.6^{>+80}_{-9.4}      $ & 0.36  & 0.47$\pm$.140 & $ 7.7^{+2.5}_{-5.5}  $ \\
P2 & 934     &     $22.2^{+85.0}_{-15.9}    $ & 0.02  & 0.28$\pm$.099 & $ 6.9^{+5.4}_{-6.5}  $ \\
P3 & 742     &     $10.0^{+61.6}_{-5.7}     $ & 0.05  & 0.24$\pm$.077 & $ 11.8^{+4.4}_{-8.1} $ \\
P4 & 550     &     $4.0^{+22.2}_{-2.5}      $ & 0.01  & 0.20$\pm$.066 & $ 14.6^{+8.9}_{-11.2}$ \\
P5 & 217     &     $8.1^{>+90}_{-4.2}       $ & 0.96  & 0.21$\pm$.069 & $ 10.4^{+6.0}_{-10.0}$ \\
    &         &                                 &       &                 &                      \\
    &         &    {\bf NW-lobe}                &       &$\rm\alpha_{inj}=0.750$&                \\
P6 & 262     &     $ 4.7^{+9.1}_{-1.9}     $ & 0.14 & 0.25$\pm$.116 & $ 14.5^{+5.9}_{-6.0} $ \\
P7 & 665     &     $ 8.4^{+87.4}_{-2.0}    $ & 2.89 & 0.34$\pm$.151 & $ 11.7^{+2.1}_{-8.2} $ \\
P8 & 857     &     $35.2^{>+90}_{-11.9}    $ & 9.73 & 0.37$\pm$.166 & $ 5.8^{+1.4}_{-5.5}  $ \\
P9 & 1049    &     $188.8^{>+85}_{-134.}   $ & 4.19 & 0.40$\pm$.175 & $ 2.5^{+2.2}_{-2.4}  $ \\
\hline
\end{tabular}
\end{table}

\begin{figure}
\vspace{148mm}
\includegraphics{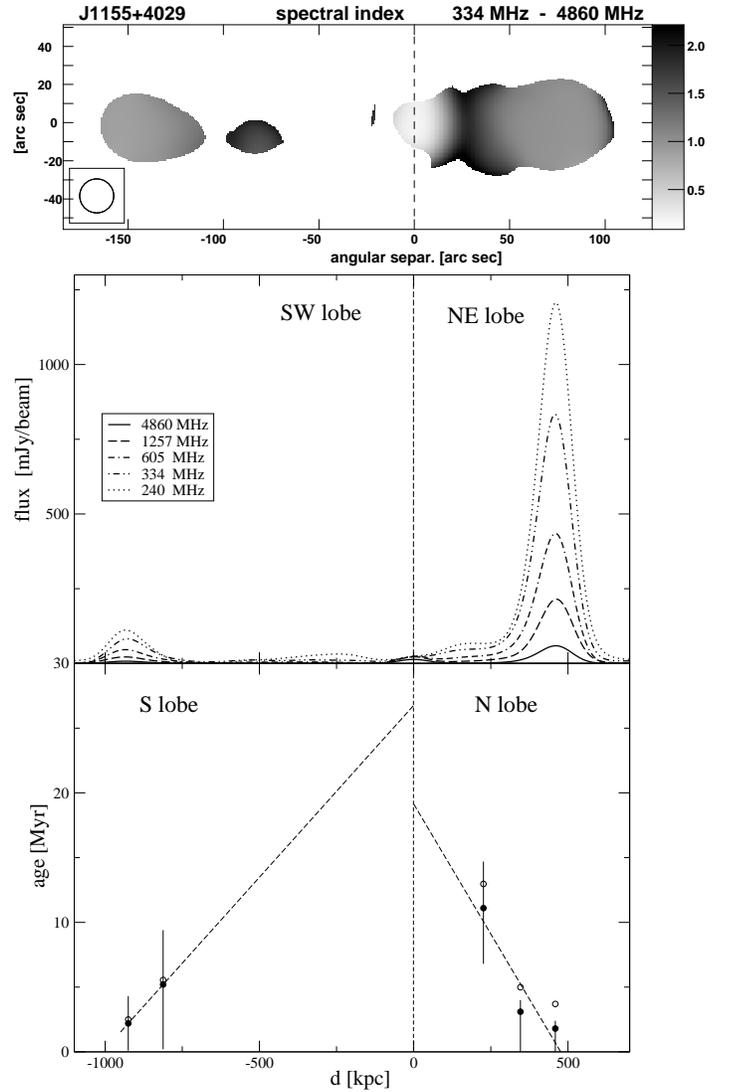}
\caption{As in Fig.\,2, but for J1155+4029}
\end{figure}

\begin{table}
\caption {As in Table 1 but for J1155+4029}
\begin{tabular}{llrlcl}
\hline
Strip & Dist. & $\nu_{br}$ & $\chi^{2}_{red}$ & B$\rm_{eq}$(rev)& $\tau_{\rm spec}$ \\
      & (kpc) &(GHz) &    & (nT) & (Myr) \\
\hline
     &         &    {\bf SW-lobe}                &       &$\rm\alpha_{inj}=0.838$&               \\
P1  & 925     &     $193.1^{+85.0}_{-80.1}$   & 0.17 & $0.84\pm.275$  & $ 2.2^{+2.1}_{4.1}$  \\
P2  & 812     &     $40.7^{>+90}_{-32.9}$     & 0.25 & $0.68\pm.231$  & $ 5.2^{+4.2}_{-5.0}$ \\
     &         &                                 &       &                &                      \\
     &         &    {\bf NE-lobe}                &       &$\rm\alpha_{inj}=0.876$&             \\
P3  & 226     &     $6.3^{+7.6}_{-2.7}$       & 0.06 & $0.93\pm.115 $ & $ 11.1^{+3.6}_{-4.3}$\\
P4  & 346     &     $51.7^{>+90}_{-20.6}$     & 1.17 & $1.23\pm.145$  & $ 3.1^{+0.9}_{-3.0}$ \\
P5  & 459     &     $81.8^{>+90}_{-34.6}$     & 1.26 & $1.62\pm.184$  & $ 1.8^{+0.6}_{-1.5}$ \\
 
\hline
\end{tabular}
\end{table}

\section{The observational results}
The results of our spectral ageing analysis for the eight GRSs selected from Paper I are
presented in Figures 2 to 9 while some of the parameters are tabulated in Tables 1 to 8.
In Figs. 2 to 9, (i) the spectral index map using
the lowest frequency GMRT data and the VLA 4.86 GHz data, (ii) the intensity profiles
of the lobes measured along the source axis at the different frequencies, and
(iii) the resulting spectral age as a function of the distance from the radio core,
are shown in the top, central and bottom panels respectively. 
Some of the numerical values are given in Tables 1 to 8 which are arranged as follows. 
Column 1: identification of the strip,
column 2: projected distance of the centre of the strip from the radio core in units of kpc,
column 3: break frequency of the spectrum of the strip according to the JP model in units of GHz,
column 4: reduced $\chi^{2}$ value of the fit,
column 5: revised magnetic-field strength in units of nT, and 
column 6: spectral age of particles in the strip.
We have estimated the physical parameters using the cosmological model where
H$_o$=71 km s$^{-1}$ Mpc$^{-1}$, $\Omega_m$=0.27, $\Omega_{vac}$=0.73 (Spergel et al. 2003).

\subsection{Notes on individual GRSs}

J0912+3510: 
there is some uncertainty as to which galaxy is associated with the radio source
(cf. Machalski et al. 2006, hereinafter referred to as MJZK).  

\begin{figure}
\vspace{143mm}
\includegraphics{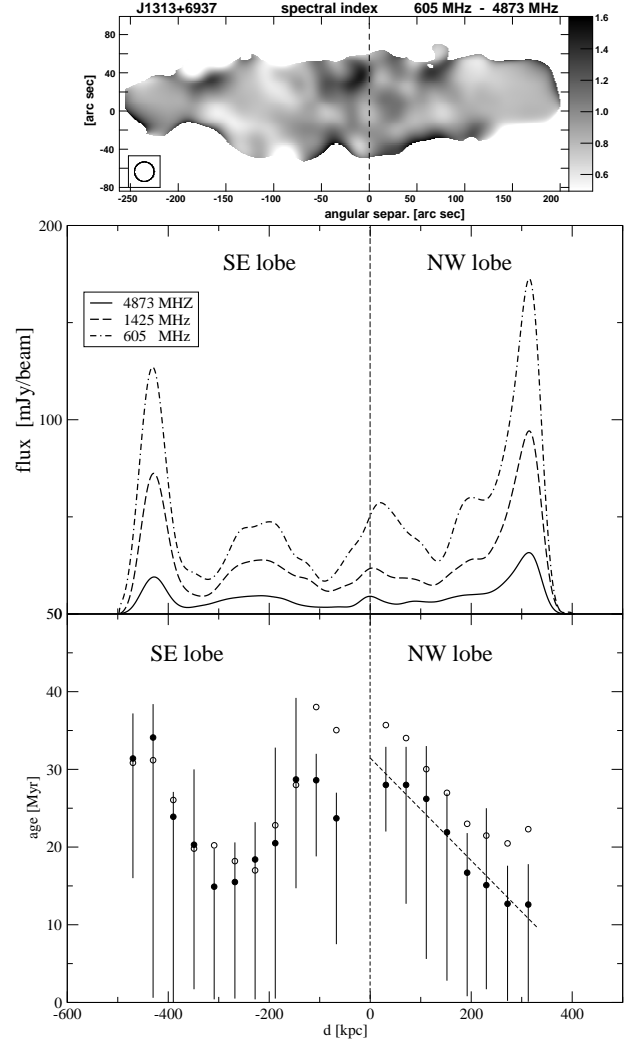}
\caption{As in Fig.\,2, but for J1313+6937}
\end{figure}

\begin{table}
\caption {As in Table 1 but for J1313+6937}
\begin{tabular}{lrrlcl}
\hline
Strip & Dist. & $\nu_{\rm br}$ & $\chi^{2}_{\rm red}$ & B$\rm_{eq}$(rev)& $\tau_{\rm spec}$ \\
      &   (kpc) & (GHz)    &                  & (nT)            & (Myr)        \\
\hline
    &          &    {\bf SE-lobe}             &        &$\rm\alpha_{inj}=0.610$&                  \\
P1 & 470      &     $8.0^{+3.6}_{-4.4}$     & 0.73  & $0.47\pm$.015 & $ 31.4^{+5.8}_{-15.4}$ \\
P2 & 430      &     $7.5^{>+80}_{-1.5}$     & 1.82  & $0.44\pm$.014 & $ 34.1^{+4.3}_{-33.5}$ \\
P3 & 390      &     $14.5^{>+90}_{-3.2}$    & 11.1  & $0.46\pm$.015 & $ 23.9^{+3.2}_{-23.5}$ \\
P4 & 349      &     $19.4^{>+80}_{-10.5}$   & 0.07  & $0.47\pm$.015 & $ 20.3^{+9.7}_{-18.6}$ \\
P5 & 309      &     $34.7^{+40.0}_{-28.9}$  & 4.38  & $0.48\pm$.015 & $ 14.9^{+4.9}_{-14.5}$ \\
P6 & 268      &     $30.9^{+34.5}_{-25.3}$  & 3.29  & $0.50\pm$.016 & $ 15.5^{+5.1}_{-15.0}$ \\
P7 & 228      &     $23.8^{+17.7}_{-18.6}$  & 3.87  & $0.47\pm$.015 & $ 18.4^{+4.8}_{-18.0}$ \\
P8 & 188      &     $22.9^{>+80}_{-17.1}$   & 0.42  & $0.40\pm$.014 & $ 20.5^{+12.3}_{-20.3}$\\
P9 & 147      &     $12.5^{+33.4}_{-5.8}$   & 0.16  & $0.38\pm$.013 & $ 28.7^{+10.5}_{-14.0}$\\
P10& 107      &     $10.1^{+2.4}_{-4.5}$    & 6.34  & $0.46\pm$.015 & $ 28.6^{+3.4}_{-9.8}$  \\
P11& 67       &     $11.3^{+3.2}_{-7.3}$    & 4.25  & $0.55\pm$.017 & $ 23.7^{+3.3}_{-16.2}$  \\ 
    &          &                              &        &                 &                      \\
    &          &    {\bf NW-lobe}             &        &$\rm\alpha_{inj}=0.610$&                 \\
P14& 31       &     $8.1^{+4.4}_{-2.2}$     & 0.38  & $0.54\pm$.017 & $ 28.0^{+4.9}_{-6.0}$  \\
P15& 71       &     $9.0^{+3.6}_{-5.2}$     & 1.22  & $0.51\pm$.016 & $ 28.0^{+4.9}_{-15.3}$ \\
P16& 111      &     $11.3^{>+80}_{-4.2}$    & 0.27  & $0.48\pm$.015 & $ 26.2^{+6.8}_{-20.6}$ \\
P17& 152      &     $15.2^{>+85}_{-5.1}$    & 1.03  & $0.50\pm$.016 & $ 21.9^{+5.0}_{-19.1}$ \\
P18& 192      &     $22.5^{>+85}_{-9.3}$    & 0.99  & $0.55\pm$.017 & $ 16.7^{+5.1}_{-15.9}$ \\
P19& 230      &     $24.9^{>+85}_{-15.9}$   & 0.001 & $0.59\pm$.018 & $ 15.1^{+9.9}_{-13.4}$ \\
P20& 272      &     $28.4^{+42.8}_{-22.5}$  & 1.55  & $0.65\pm$.019 & $ 12.7^{+4.9}_{-12.5}$ \\
P21& 313      &     $24.6^{+42.2}_{-18.8}$  & 0.96  & $0.71\pm$.020 & $ 12.6^{+5.2}_{-12.4}$ \\
\hline
\end{tabular}
\end{table}

J0927+3510: 
as mentioned in Paper I, the strips P5 and P6 in Table~2 overlap
the region of emission which was suspected to be an inner double (cf. MJZK).
However, the very steep spectra of about 1.17 and 1.35 in the strips
P5 and P6 respectively suggest that this is likely to be old lobe emission. The resulting 
spectral ages are not significantly different from the expected trend of age with
distance from the hotspot, with the age of P5 which is weaker and on the south-eastern 
side being only marginally lower.   
The somewhat higher separation speed of $\sim$0.7 and 0.4c for the eastern and western
lobe material respectively 
(Fig.\,3) suggests that the photometric-redshift estimate of 0.55 for this GRS
(MJZK) may be overestimated. 

\begin{figure}
\vspace{143mm}
\includegraphics{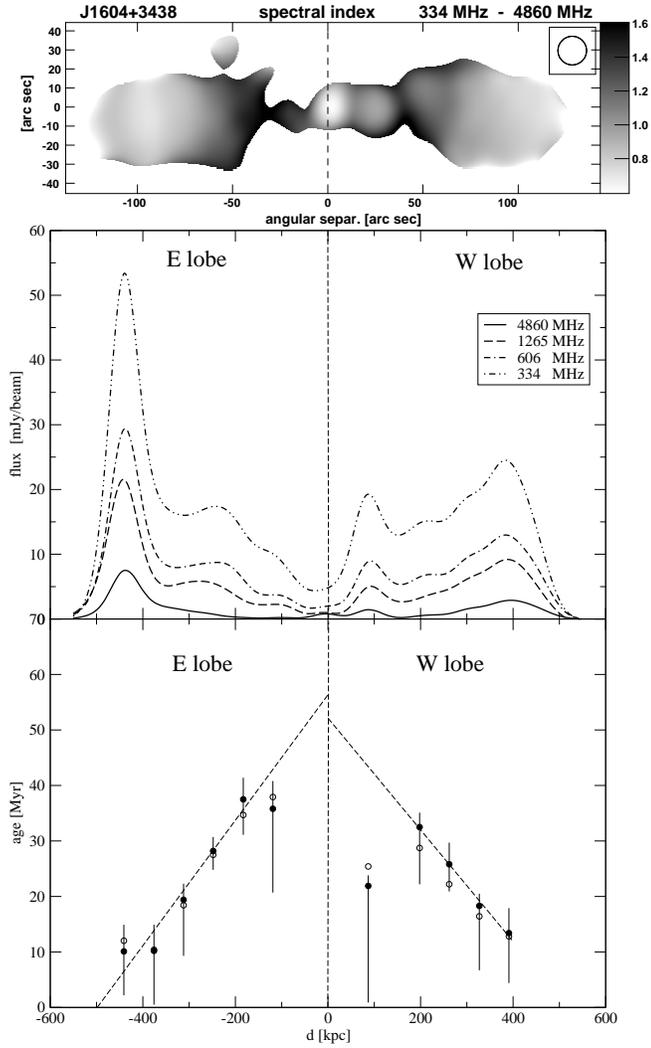}
\caption{As in Fig.\,2, but for J1604+3438}
\end{figure}

\begin{table}
\caption {As in Table 1 but for J1604+3438}
\begin{tabular}{lrrlcl}
\hline
Strip & Dist. & $\nu_{\rm br}$ & $\chi^{2}_{\rm red}$ & B$\rm_{eq}$(rev)& $\tau_{\rm spec}$ \\
      & (kpc) & (GHz)    &                  & (nT)            & (Myr)        \\
\hline
    &          &    {\bf E-lobe}             &         &$\rm\alpha_{inj}=0.554$&               \\
P1  & 441      &     $26.9^{+69.9}_{-18.1}$  &  0.82  & $0.64\pm$.021 & $10.1^{+4.8}_{-7.9}$  \\
P2  & 376      &     $33.6^{>+80}_{-17.5}$   &  0.67  & $0.52\pm$.015 & $10.2^{+4.7}_{-9.7}$  \\
P3  & 312      &     $9.6^{+3.9}_{-5.4}$     &  0.98  & $0.50\pm$.014 & $19.4^{+10.1}_{-2.9}$ \\
P4  & 248      &     $4.3^{+1.0}_{-0.8}$     &  0.16  & $0.53\pm$.016 & $28.2^{+2.5}_{-3.4}$  \\
P5  & 183      &     $2.5^{+1.2}_{-0.3}$     &  0.26  & $0.51\pm$.014 & $37.5^{+3.9}_{-6.4}$  \\
P6  & 119      &     $2.5^{+4.9}_{-0.5}$     &  1.87  & $0.57\pm$.017 & $35.8^{+5.0}_{-15.1}$ \\
    &          &                             &         &                 &                     \\
    &          &    {\bf  W-lobe}            &       &$\rm\alpha_{inj}=0.554$&                 \\
P7  & 87       &     $6.0^{>+90}_{-0.7}$     &  1.39  & $0.62\pm$.020 & $21.9^{+1.9}_{-21.0}$ \\
P8  & 198      &     $3.6^{+4.1}_{-0.5}$     &  0.93  & $0.47\pm$.013 & $32.5^{+2.6}_{-10.3}$ \\
P9  & 262      &     $6.0^{+2.3}_{-1.7}$     &  0.71  & $0.46\pm$.012 & $25.8^{+3.9}_{-4.9} $ \\
P10 & 327      &     $11.4^{+3.3}_{-7.3}$    &  1.14  & $0.47\pm$.013 & $18.3^{+11.6}_{-2.2}$ \\
P11 & 391      &     $20.2^{+25.8}_{-12.8}$  &  0.39  & $0.50\pm$.014 & $13.4^{+4.5}_{-9.0} $ \\

\hline
\end{tabular}
\end{table}

\begin{figure}
\vspace{143mm}
\includegraphics{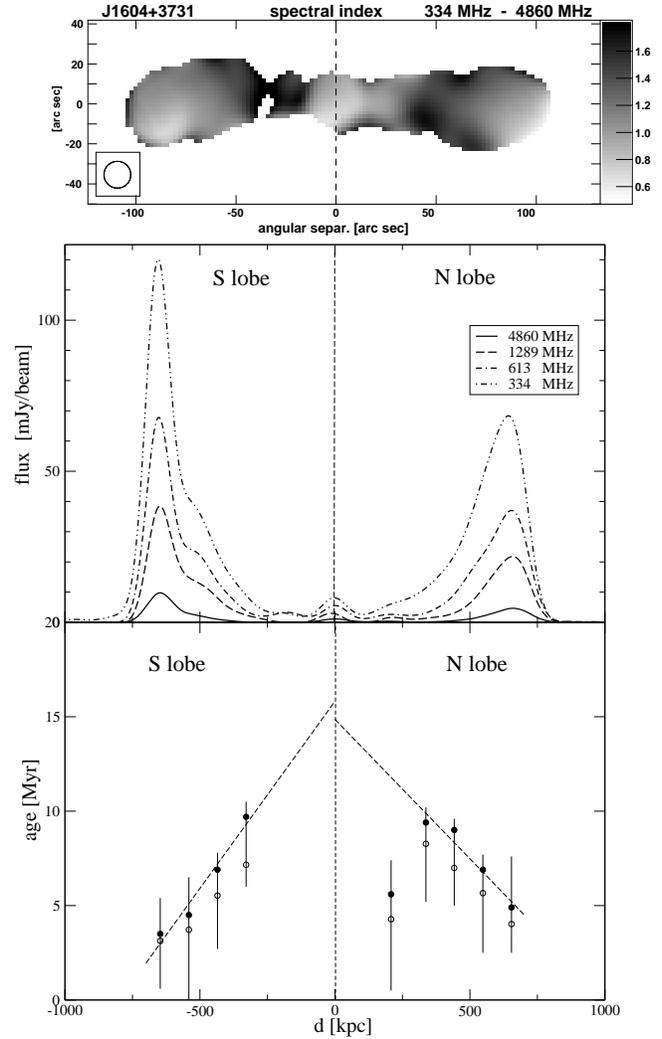}
\caption{As in Fig.\,2, but for J1604+3731}
\end{figure}

\begin{table}
\caption {As in Table 1 but for J1604+3731}
\begin{tabular}{lrrlcl}
\hline
Strip & Dist. & $\nu_{\rm br}$ & $\chi^{2}_{\rm red}$ & B$\rm_{eq}$(rev)& $\tau_{\rm spec}$ \\
      & (kpc) & (GHz)    &                  & (nT)            & (Myr)        \\
\hline
   &         &    {\bf  S-lobe}               &       &$\rm\alpha_{inj}=0.775$&              \\ 
P1 & 647     &     $31.7^{>+80}_{-18.4}$      & 0.43 & $0.78\pm$.252 & $ 3.5^{+1.9}_{-2.9}$ \\
P2 & 541     &     $20.0^{>+90}_{-10.6}$      & 3.67 & $0.70\pm$.230 & $ 4.5^{+2.0}_{-4.5}$ \\
P3 & 435     &     $8.6^{+41.8}_{-2.0}$       & 3.98 & $0.64\pm$.213 & $ 6.9^{+0.9}_{-4.2}$ \\
P4 & 329     &     $4.2^{+6.9}_{-0.6}$        & 2.14 & $0.52\pm$.179 & $ 9.7^{+0.8}_{-3.7}$ \\
   &         &                                &       &               &                      \\
   &         &    {\bf N-lobe}                &       &$\rm\alpha_{inj}=0.765$&              \\
P5 & 208     &     $ 11.1^{>+80}_{-3.8}$      & 2.79 & $0.38\pm$.141 & $ 5.6^{+1.8}_{-5.1}$ \\
P6 & 337     &     $ 4.4^{+9.7}_{-0.6}$       & 5.37 & $0.46\pm$.167 & $ 9.4^{+0.8}_{-4.2}$ \\
P7 & 442     &     $ 5.0^{+10.4}_{-0.6}$      & 2.70 & $0.56\pm$.197 & $ 9.0^{+0.6}_{-4.0}$ \\
P8 & 548     &     $ 8.5^{+48.6}_{-1.7}$      & 1.34 & $0.67\pm$.230 & $ 6.9^{+0.8}_{-4.4}$ \\
P9 & 654     &     $ 16.4^{+40.8}_{-9.5}$     & 0.90 & $0.68\pm$.234 & $ 4.9^{+2.4}_{-2.7}$ \\
\hline
\end{tabular}
\end{table}

J1155+4029: although we have tentatively identified it as a galaxy (MJZK; Paper I), 
an optical spectrum of the faint host would help resolve
the nature of the optical galaxy.  

\begin{figure}
\vspace{148mm}
\includegraphics{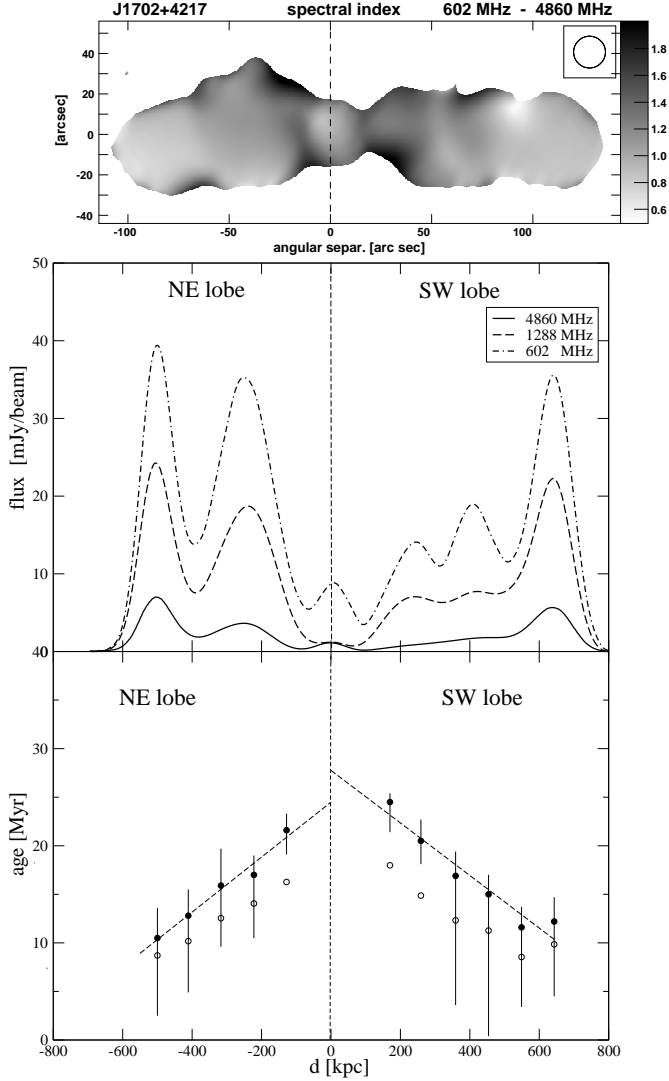}
\caption{As in Fig.\,2, but for J1702+4217}
\end{figure}

\begin{table}
\caption {As in Table 1 but for J1702+4217}
\begin{tabular}{lrrlcl}
\hline
Strip & Dist. & $\nu_{\rm br}$ & $\chi^{2}_{\rm red}$ & B$\rm_{eq}$(rev)& $\tau_{\rm spec}$ \\
      & (kpc) & (GHz)    &                  & (nT)            & (Myr)        \\
\hline
    &         &    {\bf  NE-lobe}             &       &$\rm\alpha_{inj}=0.588$&               \\ 
P1 & 500     &     $ 15.2^{+14.3}_{-10.3}$   & 1.77 & $0.49\pm$.028 & $ 10.5^{+3.2}_{-8.0}$ \\
P2 & 411     &     $ 10.5^{+61.4}_{-3.4}$    & 0.02 & $0.41\pm$.025 & $ 12.8^{+2.7}_{-7.9}$ \\
P3 & 317     &     $ 6.7^{+4.8}_{-3.2}$      & 0.29 & $0.45\pm$.026 & $ 15.9^{+3.8}_{-6.3}$ \\
P4 & 222     &     $ 5.8^{+1.5}_{-2.8}$      & 1.97 & $0.49\pm$.028 & $ 17.0^{+6.5}_{-2.0}$ \\
P5 & 128     &     $ 3.7^{+0.6}_{-0.7}$      & 2.09 & $0.40\pm$.024 & $ 21.6^{+2.5}_{-1.7}$ \\
    &         &                               &       &                &                      \\
    &         &    {\bf SW-lobe}               &       &$\rm\alpha_{inj}=0.588$&              \\
P6 & 170     &     $ 2.8^{+0.3}_{-0.5}$      & 1.65 & $0.36\pm$.022 & $ 24.5^{+3.1}_{-0.9}$ \\
P7 & 259     &     $ 4.1^{+1.1}_{-0.8}$      & 0.89 & $0.38\pm$.024 & $ 20.5^{+2.2}_{-2.4}$ \\
P8 & 359     &     $ 6.0^{>+80}_{-1.4}$      & 0.61 & $0.39\pm$.024 & $ 16.9^{+2.5}_{-13.3}$\\
P9 & 454     &     $ 7.6^{>+90}_{-1.7}$      & 1.71 & $0.39\pm$.024 & $ 15.0^{+2.0}_{-14.6}$\\
P10& 549     &     $ 12.9^{+6.2}_{-8.5}$     & 4.34 & $0.39\pm$.024 & $ 11.6^{+8.2}_{-2.1}$ \\
P11& 643     &     $ 11.3^{+6.4}_{-7.0}$     & 2.13 & $0.47\pm$.027 & $ 12.2^{+2.5}_{-7.7}$ \\

\hline
\end{tabular}
\end{table}

\begin{figure}
\vspace{150mm}
\includegraphics{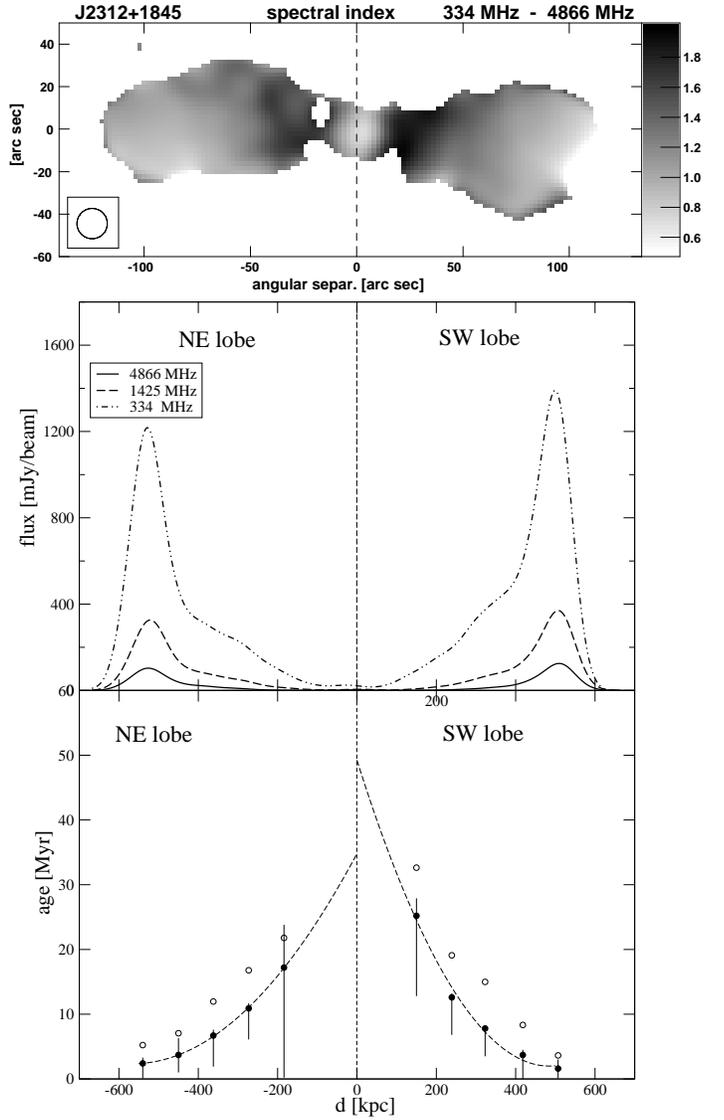}
\caption{As in Fig.\,2, but for J2312+1845/3C457}
\end{figure}

\begin{table}
\caption {As in Table 1 but for J2312+1845}
\begin{tabular}{lrrlcl}
\hline
Strip & Dist. & $\nu_{\rm br}$ & $\chi^{2}_{\rm red}$ & B$\rm_{eq}$(rev)& $\tau_{\rm spec}$ \\
      & (kpc) & (GHz)    &                  & (nT)            & (Myr)        \\
\hline
    &         &    {\bf  NE-lobe}               &       &$\rm\alpha_{inj}=0.820$&                \\ 
P1 & 540     &     $ 73.4^{>+90}_{-35.6}$     & 1.56 & $1.44\pm$.089  & $ 2.4^{+0.9}_{-2.3}$  \\
P2 & 450     &     $ 42.3^{>+80}_{-27.3}$     & 0.01 & $1.24\pm$.078  & $ 3.7^{+2.6}_{-2.7}$  \\
P3 & 362     &     $ 14.7^{>+80}_{-3.1}$      & 1.26 & $1.17\pm$.074  & $ 6.7^{+0.9}_{-4.8}$  \\
P4 & 273     &     $ 7.1^{+13.2}_{-0.9}$      & 1.45 & $1.03\pm$.067  & $ 10.9^{+0.8}_{-4.8}$ \\
P5 & 184     &     $ 4.2^{>+90}_{-4.0}$       & 12.0 & $0.82\pm$.055  & $ 17.2^{+6.6}_{-17.0}$\\
    &         &                                 &       &                &                       \\
    &         &    {\bf SW-lobe}                &       &$\rm\alpha_{inj}=0.820$&                \\
P6 & 150     &     $ 2.0^{+4.9}_{-0.4}$       & 16.6 & $0.82\pm$.055  & $ 25.2^{+2.7}_{-12.4}$\\
P7 & 239     &     $ 5.7^{+11.7}_{-0.5}$      & 11.0 & $0.99\pm$.065  & $ 12.6^{+0.6}_{-5.8}$ \\
P8 & 323     &     $ 9.2^{+30.8}_{-1.1}$      & 3.03 & $1.26\pm$.080  & $ 7.8^{+0.5}_{-4.3}$  \\
P9 & 418     &     $ 28.6^{>+90}_{-8.9}$      & 1.41 & $1.47\pm$.091  & $ 3.7^{+0.8}_{-3.7}$  \\
P10& 507     &     $ 152.6^{>+80}_{-80.0}$    & 0.10 & $1.50\pm$.092  & $ 1.6^{+1.7}_{-1.5}$  \\

\hline
\end{tabular}
\end{table}

J1313+6937, DA\,340: the spectral-index map and the intensity profiles
along the source axis suggest that the synchrotron radiation observed in separate
strips of this GRS may
be related to a mixture of emitting particles which were injected or accelerated
at different time epochs. The SE lobe of this GRS is the only one, among the other
lobes studied in this Paper, in which the synchrotron ages determined do not more
or less gradually increase with distance from the hotspot area to the core. 
The spectral
ageing analysis of the entire lobes of this galaxy was published by
Schoenmakers et al. (2000) where the radio spectrum of each of the lobes were
determined within the frequency range of 325 MHz--10.5 GHz. Though their $\alpha_{\rm inj}$
values found fitting the JP model, i.e. 0.71$\pm$0.02 and 0.75$\pm$0.02 for the SE
and NW lobes, respectively, differ from our values, their average synchrotron ages of
these lobes of 28$\pm$1 Myr and 27$\pm$1 Myr are consistent with our results
(cf. Table 4).

J1604+3438: 
the inner part of the structure comprising the flat-spectrum radio
core and the two nearby peaks of emission seen most clearly in the image
published by MJZK led them to classify the source as a possible double-double radio 
galaxy (DDRG).  However as noted in Paper I spectral indices of these features 
are steep and one needs to image them with higher resolution in both 
total intensity and linear polarisation to investigate the DDRG
nature of the source. It is relevent to note that the spectral age of the
western feature lies significantly below the age$-$distance relationship 
for the western lobe.  

J1604+3731: 
this is the highest-redshift radio galaxy in the sample. 
The relatively younger age seen in the strip closest to the core in the 
northern lobe is coincident with a knot or peak of emission (Konar et al.
2004; Paper I).

J1702+4217:
although there are peaks of emission visible in the lobe emission, especially at
the lower frequency, the spectral age increases smoothly with distance from the hotspots.

J2312+1845, 3C457:
the only source in the sample showing undoubtedly a curvature in the age$-$distance plot.
This could be due to deceleration of the radial
expansion velocity of the lobe material. The data in Table 8 give an expansion
velocity of about 0.11$c\pm$0.05$c$ in the vicinity of the leading heads which slows
to about 0.03$c\pm$0.015$c$ in the regions where the oldest particles are detected. 
Such curvatures may also be produced by re-acceleration of particles in different
regions of the lobes. 

\subsection{Separation speeds and spectral ages}
The plots in the bottom panels of Figs.\,2--9 show that, in general, the spectral age of 
the emitting particles increases systematically with distance from the leading heads of the lobes, 
i.e the hotspot regions. These plots have been shown using the magnetic field strengths
estimated using the BK formalism  (full circles) 
as well as the classical-1 (Miley 1980; see Paper I) minimum-energy one (open circles). 
These two field strengths
differ by a factor of $\sim$3 (cf. Paper I). The numerical values we list 
here have been estimated using the BK formalism. 
For most of these lobes we calculate the linear regression of the ages 
($\tau_{\rm spec}$) on the corresponding distances ($d$), which is shown as a dashed line
in Figs.\,2--9. The slopes of these lines give a characteristic speed, $v_{\rm sep}$,
which is an indication of the  average speed of the lobe material relative to the hotspots.
A clear exception is J2312+1845 (3C457) for which the speed,  $v_{\rm sep}$, evidently 
decelerates with distance from hotspot regions. This source is much stronger than the remaining 
GRSs, therefore the errors of the spectral break frequency, $\rm \nu_{br}$, and the corresponding
spectral age in each of the strips are much smaller than those in our remaining sources. 
In other cases the linear regression fit is the most conservative, though there is no physically
justified reason for the material separation speed to be constant.  

Our estimates of the average speed and a `characteristic' spectral age for the lobes
in the regions of the observed emission from the lobes as well as the values when
extrapolated to the positions of the cores are presented in Table 9. These values 
have been presented for all the sources except for the SE-lobe of J1313+6937 which 
has a complex age$-$distance plot. In Table 9 the values have been presented for the
estimates using the magnetic field values from both the BK and the classical-1
formalisms. The Table is arranged as follows. Column 1: source name; column 2: the optical 
identification; column 3: redshift; column 4: largest linear size $l$ in kpc; column 5: log of
the luminosity in units of W Hz$^{-1}$; column 6: lobe identification; column 7: spectral age
using the BK formalism, denoted by `rev', till the most distant strip from the hotspot; column 8: spectral
age when the age-distance relationship is extrapolated to the region of core; columns 9 and 10: same
as columns 7 and 8 but using the classical-1 estimates of the magnetic field, denoted by `class'; 
columns 11 and 12: estimates of the separation velocity for the BK and classical-1 approaches.

\begin{table*}
\begin{center}
\caption{Some of the physical parameters of the sources}
\begin{tabular}{l c l r c  l cc cc cc}
\hline
Source   & Opt. & red-  &  $l$  &P$_{1.4}$  &Lobe& age  & ext-age& age & ext-age &  $v_{\rm sep}/c$ & $v_{\rm sep}/c$ \\
         & Id.  & shift &       &           &    &  rev &  rev   & class & class  &    rev    & class          \\
         &      &       & (kpc) &(W Hz$^{-1}$)&  & (Myr)& (Myr) & (Myr) & (Myr)  &           &              \\  
(1)      & (2)  & (3)   & (4)   & (5)       & (6) &(7)  & (8)   &  (9)  & (10)   &   (11)    &   (12)      \\
\hline                                                    

J0912+3510 & G & 0.2489 & 1449 &  25.44     & N  &28.7  &49.9    &23.0   &34.2    &   0.09           & 0.18         \\
           &   &        &      &            & S  &32.6  &35.1    &22.2   &23.2    &   0.07           & 0.12         \\
J0927+3510 & G & 0.55$^\ast$ & 2206 & 26.02 & SE &14.6  &21.4    &8.3    &10.4    &   0.24           & 0.70        \\
           &   &        &      &            & NW &14.5  &19.6    &8.7    &12.2    &   0.21           & 0.37        \\
J1155+4029 & G & 0.53$^\ast$ & 1437 & 26.55 & SW &5.2   &26.8    &5.6    &27.5    &   0.12           & 0.12         \\
           &   &        &      &            & NE &11.1  &19.1    &13.0   &21.0    &   0.08           & 0.08         \\
J1313+6937 & G & 0.106  &  745  & 25.60     & SE &23.7  & --     &35.1   & --     &   --             & --           \\
           &   &        &      &            & NW &28.0  &31.5    &35.7   &36.4    &   0.05           & 0.06         \\
J1343+3758$^{1}$ & G & 0.2267 & 2463 & 25.32& NE &33.5  &60.8    &26.5   &53.4    &   0.13           & 0.10         \\
           &   &        &      &            & SW &36.2  &52.5    &33.6   &50.0    &   0.12           & 0.09         \\
J1453+3308$^{2}$ & G & 0.249 & 1297 & 25.93 & S  &58.4  &65.0    &47.9   &61.8    &   0.04           & 0.05         \\
           &   &        &      &            & N  &46.6  &54.9    &38.8   &53.1    &   0.04           & 0.05          \\
J1604+3438 & G & 0.2817 & 846 & 25.53       & E  &56.3  &56.3    &34.7   &50.8    &   0.03           & 0.03         \\
           &   &        &      &            & W  &52.1  &52.1    &28.7   &44.5    &   0.03           & 0.04         \\
J1604+3731 & G & 0.814  & 1346 & 26.60      & S  &15.8  &15.8    &7.2    &11.3    &   0.16           & 0.25         \\
           &   &        &      &            & N  &14.9  &14.9    &8.3    &12.8    &   0.22           & 0.24         \\
J1702+4217 & G & 0.476  & 1160 & 26.19      & NE &24.5  &24.5    &16.3   &18.8    &   0.12           & 0.16         \\
           &   &        &      &            & SW &27.8  &27.8    &18.0   &19.9    &   0.12           & 0.18         \\
J2312+1845 & G & 0.427 & 1056 & 27.11       & NE &34.8  &34.8    &21.9   &36.5    &   0.08$^3$       & 0.07$^3$     \\
           &   &        &      &            & SW &49.2  &49.2    &32.6   &54.1    &   0.05$^3$       & 0.04$^3$     \\
\hline
\end{tabular}

Notes:
$^\ast$ denotes an estimated redshift;
(1) calculated from the data in Jamrozy et al. (2005);
(2) calculated from the data in Konar et al. (2006);
(3) speed estimated using the two extreme strips for each lobe.

\end{center}
\end{table*}

\begin{figure}
\psfig{file=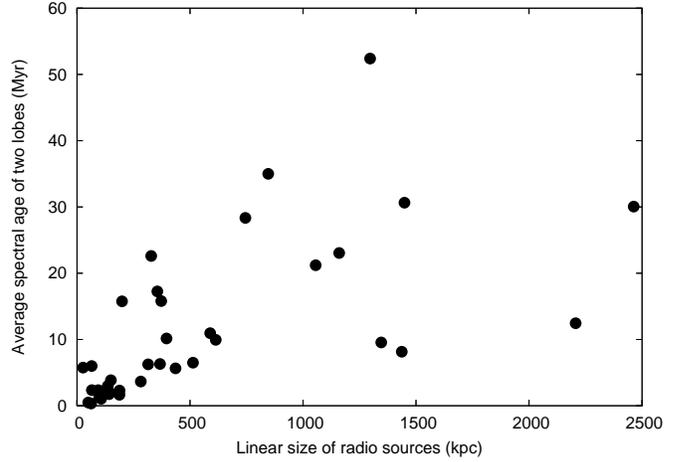,width=3.5in,angle=-90}
\caption{The spectral age in Myr as a function of the largest linear size in kpc}
\end{figure}

\begin{figure}
\vspace{50mm}
\includegraphics{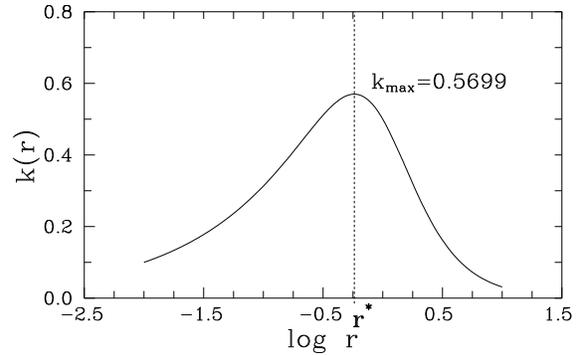}
\caption{Function $k(r)$}
\end{figure}

\section{Discussion}
\subsection{Spectral ages}
The maximum spectral ages estimated for the detected radio emission in the lobes
of our sources range from $\sim$6 to 36 Myr with a median value of $\sim$20 Myr
using the classical equipartition fields. Using the magnetic field estimates from
the BK formalism the spectral ages range from $\sim$5 to 38 Myr with
a median value of $\sim$22 Myr. These ages are significantly older than smaller sources.
The median linear size of these large sources is $\sim$1300 kpc. For comparison, we 
first consider the sample
of smaller sources studied by Leahy et al. (1989) which has been 
observed over a large frequency range. They estimated the spectral ages of
a sample of 16 3CR radio sources from low-frequency observations with 
the Multi-Element Radio Linked Interferometer Network (MERLIN) at 151 MHz and
high-frequency observations at 1500 MHz with the VLA. We have re-estimated 
their values using the cosmological parameters used in this paper. The ages of the lobes
of these sources range from $\sim$2.5 to 26 Myr with a median value of $\sim$8 Myr,
significantly smaller than the value for the GRSs. The median value of the 
linear size for this sample of sources is 342 kpc. 

Considering the sample of 14 high-luminosity double radio sources studied
by Liu et al. (1992), we have again re-estimated the spectral ages using the
cosmological parameters used in this paper, 
and find that the values range from $\sim$0.3 to 5.3 Myr with a median value of 
$\sim$1.7 Myr.  These sources have even smaller spectral ages and their 
median linear size is 103 kpc. Considering these different samples, there is
a trend for the spectral ages to increase with linear size (Fig. 10) as has been
noted earlier in the literature (e.g. Parma et al. 1999; Murgia et al. 1999; Murgia 2003). 
It is also interesting to
note that the relative speeds of the lobe material has been estimated to be
in the range of $\sim$0.03 to 0.2c with a median value of $\sim$0.1c for the
Liu et al. sample. In comparison, the corresponding speeds for our sample of
GRSs is about 0.03 to  0.25 (excluding J0927+3510 for which the photometric
redshift value may be overestimated, cf. Sect.~3.1), with a median value of 
$\sim$0.09c, similar to that of the  the Liu et al. sample.

The above trend for spectral age to increase with size is broadly consistent with 
the expectations of dynamical models of the propagation of jets in an external
medium (e.g. Falle 1991; Kaiser \& Alexander 1997; Jeyakumar et al. 2005 and 
references therein) in which the linear size of the source is a function of its 
(dynamical) age. 
Nevertheless, while interpreting these numbers caveats related to the evolution of
the local magnetic field in the lobes need to be borne in mind (e.g. Rudnick, Katz-Stone
\& Anderson 1994; Jones, Ryu \& Engel 1999; Blundell \& Rawlings 2000).
Also, while Kaiser (2000)
have suggested that spectral and dynamical ages are comparable if
bulk backflow and both radiative and adiabatic losses are taken into account in a
self-consistent manner, Blundell \& Rawlings (2000) suggest that this may be so only
in the young sources with ages much less than 10 Myr. In the study of the FRII type
giant radio galaxy, J1343+3758, Jamrozy et al. (2005) find the dynamical age to be
approximately 4 times the maximum synchrotron age of the emitting particles.

In the plots of spectral age against distance from the hotspots, we note that
none of the sources show zero age at the hotspots. In fact the ages in the hotspots
usually range from a few to $\sim$20 Myr, except in the case of the south-eastern lobe
of J1313+6937 where the age in the outer extremity is $\sim$30 Myr. This tendency for 
the hotspots to exhibit a non-zero age has been noted earlier by a number of authors 
(e.g. Liu et al. 1992). In the Liu et al. sample, ages in the hotspots are typically
less than $\sim$1 Myr while in the case of Cygnus A, Carilli et al. (1991) find the
ages of the hotspots to be 0.1 and 0.2 Myr for the north-western and south-eastern lobes
respectively. The typical linear resolution of the sources observed by Liu et al. is 
$\sim$10 kpc for most sources while that of Cygnus A is $\sim$5 kpc. In comparison, the linear
resolution of our GRSs ranges from $\sim$30 to 110 kpc with a median value of $\sim$80 kpc.
For comparison, the sizes of hotspots range from about a few parsec to $\sim$10 kpc increasing 
with source size till the largest linear size is $\sim$100 kpc and then flattens for larger
source sizes  (e.g. Jeyakumar \& Saikia 2000). Possible explanations for 
the non-zero ages include (i) the possibility that the hotspots have been inactive for 
10$^5$ to 10$^6$ yr, (ii) contamination by more extended emission and (iii) higher 
magnetic fields in the hotspots (cf. Liu et al.). The coarse resolution of our observations
suggest that contamination by extended emission is the likely cause for the non-zero
ages in the hotspots of our GRSs.   
  
Another effect, visible in the bottom panels of  Figs.\,2--9 and related to the magnetic fields, 
is the age difference, $\Delta=\tau_{\rm spec}(\rm rev)-\tau_{\rm spec}(\rm class)$, which is
either positive or negative. This is a simple consequence of Eq.\,(1). Given
$\nu_{\rm br}$ and $z$ imply the maximum $\tau_{\rm spec}$ for $B$=$B_{\rm iC}/\sqrt{3}$.
Substitution of the ratio $r\equiv B/B_{\rm iC}$ into Eq.\,1 gives

\[\tau_{\rm spec}=50.3\frac{r^{1/2}}{1+r^{2}}B_{\rm iC}^{-3/2}\{\nu_{\rm br}(1+z)\}^{-1/2}=\]
\[=50.3\,k(r)\{0.318(1+z)^{2}\}^{-3/2}\{\nu_{\rm br}(1+z)\}^{-1/2}.\]

\noindent
The function $k(r)$ has a maximum at $r^{*}$=$3^{-1/2}$, thus the maximum of
$\tau_{\rm spec}$ can be written as

\[\tau_{\rm spec}^{\rm (max)}=159.8\,\nu_{\rm br}^{-1/2}(1+z)^{-7/2}\]
 
\noindent
The function $k(r)$ as a function of $r$ is plotted in Fig.\,11.
 
In our calculations the values of the magnetic field calculated with the revised
formula, $B_{\rm eq}$(rev), are always greater than the values of $B_{\rm eq}$(class)
calculated with the classical formula of Miley (1980) where, for a comparison with
several previously published papers on spectral ages of other radio sources, we have
assumed the proton to electron energy ratio to be unity.
Therefore, if both values of $B_{\rm eq}$(class) and $B_{\rm eq}$(rev) are less 
than $B_{\rm iC}/\sqrt{3}$ (or $r<r^{*}$; see Fig.\,11), then the ages derived with
the classical formula (open circles in Figs.\,2--9) are lower than those derived
with the revised formula (full circles in Figs.\,2--9). On the contrary, if both values
of the magnetic field are either close to, or greater than  $B_{\rm iC}/\sqrt{3}$,
the `classical' ages are greater than the `revised' ages. 

\subsection{Injection spectral indices}
The injection spectral indices which have been estimated from fits to the spectra
of the lobes or the entire source using our measurements as well as low-frequency
flux density values at 151 MHz from the Cambridge surveys (Green 2002 and references
therein) and the 
VLA Low-frequency Sky Survey (VLSS; Cohen et al. 2007; {\tt http://lwa.nrl.navy.mil/VLSS})
estimates for the stronger sources. $\alpha_{\rm inj}$ varies from $\sim$0.55 to 0.88
with a median value of $\sim$0.6 for our sources.  For strong, non-relativistic shock in a 
Newtonian fluid $\alpha_{\rm inj}$ = 0.5 (Bell 1978a,b; Blandford \& Ostriker 1978).
However, for relativistic shocks or shocks in which  fields and relativistic particles
alter the shock dynamics, values of $\alpha_{\rm inj}$ are in the range of 0.35
to 0.65 (Heavens 1989; Kirk \& Schneider 1987; Drury \& Volk 1981; Axford, Leer
\& McKenzie 1982). Carilli et al. (1991) estimate $\alpha_{\rm inj}$ to be 0.5
for Cygnus A while Meisenheimer et al. (1989) obtain similar values for a few
powerful radio galaxies. For the sample of sources studied by Liu et al. (1992),
the values of $\alpha_{\rm inj}$ have been estimated from the low-frequency spectral
indices using data between 38 MHz and 1 GHz. Their values of $\alpha_{\rm inj}$
range from 0.65 to 1 with a median value of 0.77. The values of the spectral indices
of the hotspots estimated from low-frequency measurements by Leahy et al.
(1989), which is similar to $\alpha_{\rm inj}$, ranges from 0.64 to 1.16 with a median
value of $\sim$0.8. Katz-Stone \& Rudnick (1997) also estimate a steep injection 
spectral index of 0.8 for the compact steep spectrum source 3C190.  
It is important to determine the spectra of the lobes over a wide 
low-frequency range using the GMRT and upcoming instruments such as the Long
Wavelength Array (LWA) and the Low Frequency Array (LOFAR) to get better estimates of the 
injection spectral indices, and also compare these with estimates from the inverse-Compton 
scattered x-ray spectrum (e.g. Harris 2005). 

\begin{figure}
\vbox{
\psfig{file=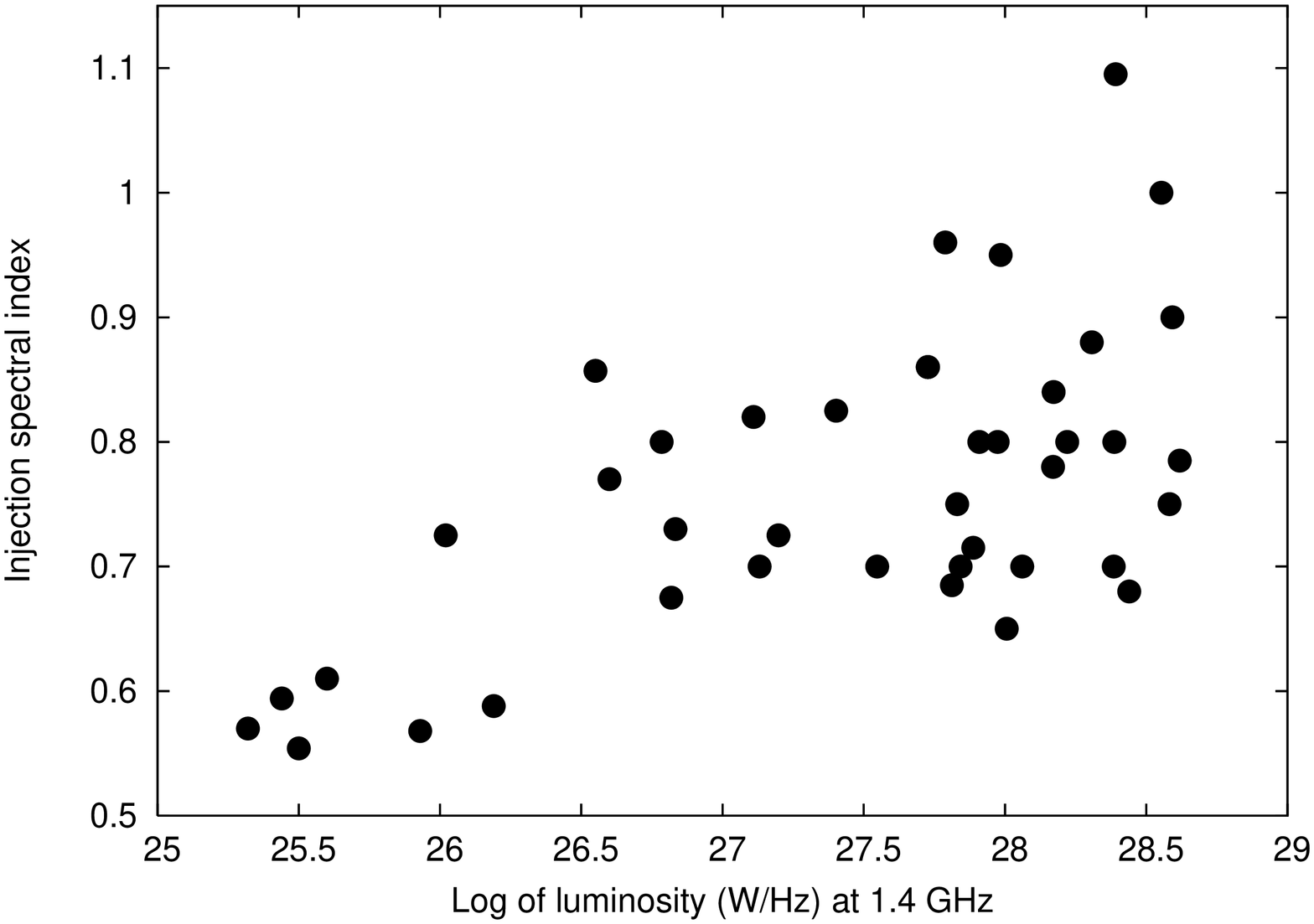,width=3.5in,angle=-90}
\psfig{file=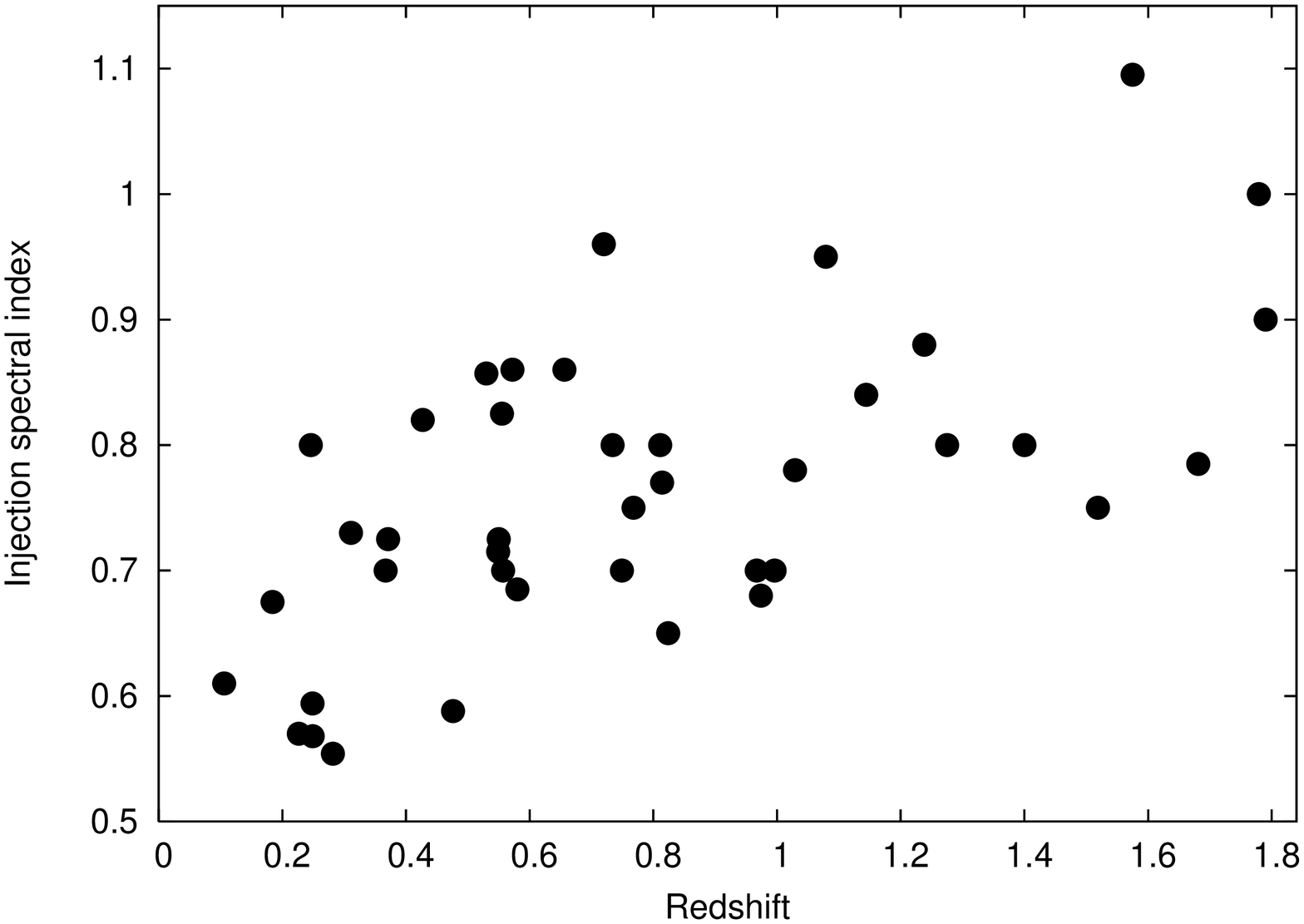,width=3.5in,angle=-90}
     }
\caption{The injection spectral index, $\alpha_{\rm inj}$, as a function of radio luminosity at
         1.4 GHz in units of W Hz$^{-1}$ (upper panel) and redshift (lower panel).}
\end{figure}

\begin{figure}
\vbox{
\psfig{file=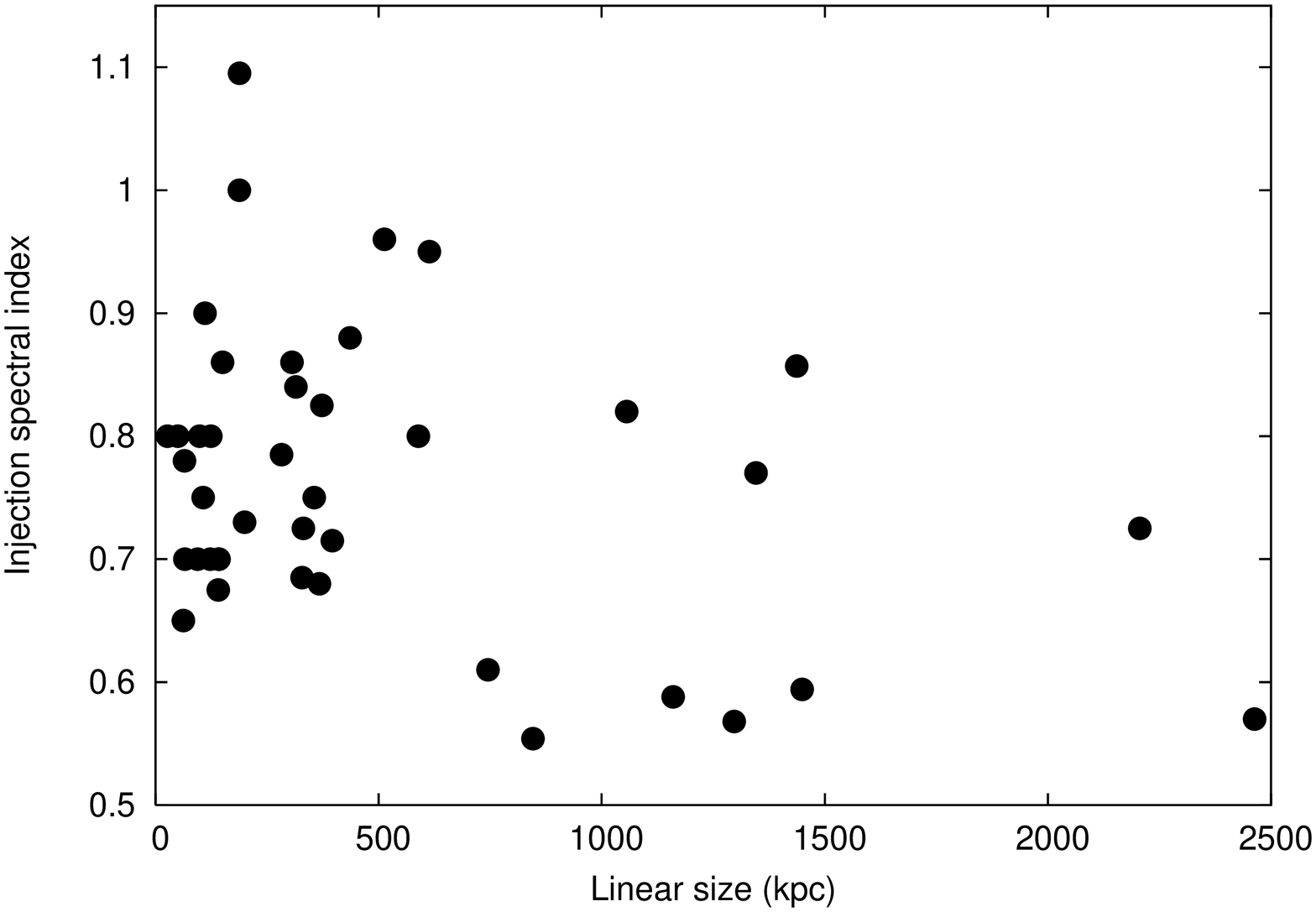,width=3.5in,angle=-90}
\psfig{file=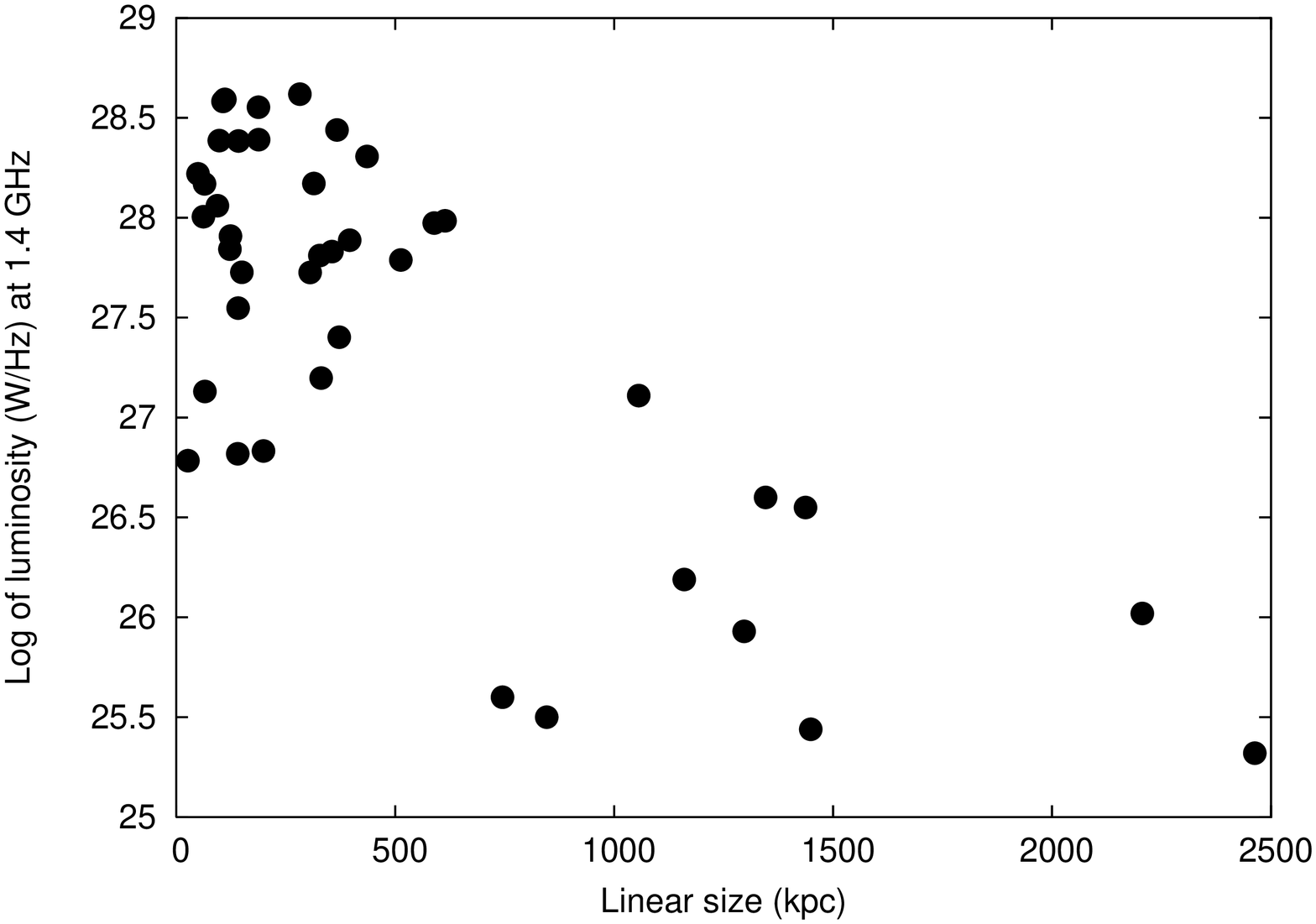,width=3.5in,angle=-90}
     }
\caption{The injection spectral index, $\alpha_{\rm inj}$, as a function of the largest linear size in kpc
       (upper panel) and the luminosity$-$linear size diagram for the sample of sources (lower panel). }
\end{figure}

We have investigated any possible dependence of $\alpha_{\rm inj}$ on source size using 
our estimates and those of Liu et al. (1992) and Leahy et al. (1989). We find 
$\alpha_{\rm inj}$ to be strongly correlated with luminosity (Fig.~12, upper panel), consistent with the well-known 
spectral-index luminosity correlation  for extragalactic radio sources (e.g. Laing \& Peacock
1980, and references therein).  
However, lower frequency measurements than those we have used using instruments such as the GMRT,
LOFAR and LWA would be extremely useful to better determine the values of $\alpha_{\rm inj}$.  
$\alpha_{\rm inj}$ also appears to be strongly correlated 
with redshift (Fig.~12, lower panel), but since most of our sources come from a reasonably narrow range 
of flux density it is difficult to distinguish whether the fundamental correlation is with luminosity
or redshift. It is worth noting that while exploring correlations of $\alpha_{\rm inj}$ with
either luminosity or redshift, K-correction factors due to cosmological redshifts should not
play a significant role. A plausible scenario for such a correlation could be multiple shocks
in the heads of the jets as suggested by Gopal-Krishna \& Wiita (1990).  
We have also explored possible correlation  of $\alpha_{\rm inj}$ with source size and find that
our estimates for the GRSs are smaller than the corresponding values for the smaller sources 
from the Liu et al. and Leahy et al. samples, which tend to have more prominent hotspots (Fig. 13, upper panel). 
The inverse correlation 
of luminosity and size for these samples (Fig. 13, lower panel) which define the approximate 
upper envelop of the luminosity-size diagram (e.g. Kaiser, Dennett-Thorpe \& Alexander 1997; Blundell, 
Rawlings \& Willott 1999; Ishwara-Chandra \& Saikia 1999) are consistent with the above trends.   

\section{Concluding remarks}

We summarise briefly the main results of our analysis.

(i)  The maximum spectral ages estimated for the detected radio emission in the lobes
     of our sources range from $\sim$6 to 36 Myr with a median value of $\sim$20 Myr
     using the classical equipartition fields. Using the magnetic field estimates from
     the BK formalism the spectral ages range from $\sim$5 to 38 Myr with
     a median value of $\sim$22 Myr. These ages are significantly older than those of smaller 
     sources (e.g. Leahy et al. 1989;  Liu et al. 1992). The GRSs are older sources possibly 
     evolving in underdense regions of the intergalactic medium 
     (cf. Cotter 1998; Mack et al. 1998; Machalski, Chy\.{z}y \& Jamrozy 2004), 
     although the external medium is often asymmetric on scales of $\sim$1 Mpc
     (e.g. Ishwara-Chandra \& Saikia 1999; Konar et al. 2004, 2007). 
        
     The spectral ages depend on the estimated magnetic fields. Strengths of
      the equipartition field calculated with the revised formalism proposed by BK
      are greater than the corresponding values provided by the classical
      formula of Miley (1980) by a factor of $\sim$3 (Paper I). However, the 
      inferred synchrotron age is a function of the ratio
      $B/B_{\rm iC}$. If $B$ is greater than $B_{\rm iC}/\sqrt{3}$, the `revised' age is lower
      than the `classical' age. 

(ii) The injection spectral indices range from 0.55 to 0.88 with a median value of
      $\sim$0.6. Our estimates for the GRSs are marginally smaller than those estimated for
      smaller sources by Leahy et al. (1989) and  Liu et al. (1992). Reliable
      low-frequency measurements of the lobes using instruments such as GMRT, LWA and LOFAR are
      required to get more reliable  estimates of the injection spectral indices. We have
      explored possible correlations of $\alpha_{\rm inj}$ with other physical parameters
      and find that it appears to increase with luminosity and redshift, but shows an
      inverse correlation with linear size. 

\section*{Acknowledgments}
We thank our reviewer Ruth Daly, and 
Gopal-Krishna, Vasant Kulkarni and Paul Wiita for their comments on the manuscript.
The Giant Metrewave Radio Telescope (GMRT) is a national facility operated by the
National Centre for Radio Astrophysics (NCRA) of the Tata Institute of Fundamental Research.
We acknowledge access to the SYNAGE software kindly provided by Dr K.-H. Mack
and Dr M. Murgia (Istituto di Radioastronomia, Bologna, Italy). MJ is indebted to
Professor D.J. Saikia and the staff members for hospitality during his stay at the NCRA.
JM and MJ acknowledge the MNiSW funds for scientific research in years 2005--2008 under
contract No. 0425/PO3/2005/29.

\end{document}